\documentclass[journal,final,twoside]{IEEEtran}

\usepackage{balance}
\usepackage{cite}
\usepackage{graphicx}
\usepackage[cmex10]{amsmath}
\usepackage{amsfonts}
\usepackage{array}
\usepackage{mdwmath}
\usepackage{mdwtab}
\usepackage[tight,footnotesize]{subfigure}
\usepackage{multirow}
\usepackage{amssymb}
\usepackage{dsfont}
\usepackage{rotating}

\usepackage[noend]{algpseudocode}
\usepackage{pdfpages}

\newcommand*{\MyPath}{}

\algnewcommand\algorithmicinput{\textbf{Input:}}
\algnewcommand\Input{\item[\algorithmicinput]}
\algnewcommand\algorithmicoutput{\textbf{Output:}}
\algnewcommand\Output{\item[\algorithmicoutput]}
\renewcommand{\theenumi}{\alph{enumi}\)}
\renewcommand{\labelenumi}{\theenumi}

\renewcommand{\(}{\ifmmode\left(\else{(}\fi}
\renewcommand{\)}{\ifmmode\right)\else{)}\fi}
\newcommand{\lv}{\left\lvert}
\newcommand{\rv}{\right\rvert}

\newlength{\origCol}
\newtheorem{theorem}{Theorem}
\newtheorem{definition}[theorem]{Definition}

\newtheorem{proposition}[theorem]{Proposition}

\newtheorem{lemma}[theorem]{Lemma}

\definecolor{darkblue}{rgb}{0,0,.5}

\newcommand{\RM}{\text{RM}}

\newcommand{\Chen}{\text{Chen}}
\newcommand{\New}{\text{New}}

\usepackage{url}

\usepackage{flushend}

\begin{document}

\title{An Improved Majority-Logic Decoder \\ Offering Massively Parallel Decoding for \\ Real-Time Control in Embedded Systems}

\author{Juliane Bertram, Peter Hauck, and Michael Huber,~\IEEEmembership{Member,~IEEE}\\
\thanks{%
Manuscript received February 6, 2013; revised July 20, 2013. The editor coordinating the
review of this paper and approving it for publication was T.-K. Truong.}
\thanks{
The work of M.~Huber was supported by the Deutsche Forschungsgemeinschaft \(DFG\) via a Heisenberg grant \(Hu954/4\) and a Heinz Maier-Leibnitz Prize grant \(Hu954/5\).
The work of P.~Hauck was supported by Proyecto MTM2010-19938-C03-02, Ministerio de Ciencia e Innovaci\'{o}n, Spain.}
\thanks{The authors are with the Department of Computer Science, Eberhard Karls Universit\"at T\"ubingen, Sand~13,
D-72076 T\"ubingen, Germany (e-mail: \{bertramj, hauck, huber\}@informatik.uni-tuebingen.de).}
\thanks{%
Digital Object Identifier 10.1109/TCOMM.2013.13.130109}}

\maketitle

\markboth{IEEE Transactions on Communications, accepted for publication}{Bertram \MakeLowercase{\textit{et al.}}: An Improved Majority-Logic Decoder Offering Massively Parallel Decoding for Real-Time Control $\ldots$}

\pubid{0090-6778/13\$31.00~\copyright~2013 IEEE}

\pubidadjcol

\begin{abstract}
We propose an easy-to-implement hard-decision majority-logic decoding algorithm for Reed--Muller codes $\RM\(r,m\)$ with $m \geq 3,$ $m/2 \geq r\geq 1.$ The presented algorithm outperforms the best known majority-logic decoding algorithms and offers highly parallel decoding. The result is of special importance for safety- and time-critical applications in embedded systems. A simple combinational circuit can perform the proposed decoding. In particular, we show how our decoder for the three-error-correcting code $\RM(2, 5)$ of dimension 16 and length 32 can be realized on hardware level.
\end{abstract}


\section{Introduction}
\PARstart{E}{mbedded} systems are becoming ubiquitous and an integral part of our everyday life. Addressing functional safety is a major challenge with increasing complexity. Typical examples of safety-critical embedded systems include vehicle safety or driver assistance systems with accident prevention. However, functional safety is becoming more prevalent not just in the automotive sector, but also in industrial markets such as aviation, solar energy, and the medical sector \(e.g.,~\cite{Freescale11}\). Memory devices increasingly provide built-in error correction in order to restore corrupted data~\cite{Issi11} and also to maximize the number of writes in flash memory~\cite{Huang11}. With information and communication technology components becoming ever smaller and more complex, the probability for hardware immanent error arises.

Decoders taking advantage of the cyclic structure of shortened Reed--Muller codes accommodate the increasing demand for less space consumption -- at the cost of the decoding duration~\cite{Blahut03}. On the other hand, several recursive algorithms were developed allowing decoding with only $O\(\min\(r,m-r\)\cdot 2^m\)$ operations~\cite{Dumer04,SchnBos95}. Though the number of operations could be reduced,  all these operations need to be executed one after another. Therefore, these algorithms require much parallel time. Parallel time is defined as the time the algorithm takes if all its modules are parallelized to the maximum possible amount. Thus, cyclic as well as recursive decoders are not designed for correcting errors in parallel. However, for all safety-critical applications, where real-time control is ranked first, decoding multiple positions in parallel saves precious time. Decoders based on majority-logic can accomplish this task. Furthermore, in embedded systems, very simple hard-decision algorithms are mostly preferable to soft-decision algorithms~\cite{Dumer04,SchnBos95}. Therefore, hard-decision decoders for Reed--Muller codes using decision by majority are an attractive option for forward error correction in real-time on hardware level.

\pubidadjcol

A majority-logic decoding algorithm was first proposed by Reed~\cite{Reed54}. Reed's algorithm consists of $r+1$ decoding steps in which majority voting is performed. Chen~\cite{Chen71,Chen72} significantly improved Reed's decoding algorithm by reducing the number of decoding steps. In particular, if Reed--Muller codes $\RM\(r,m\)$ with $m \geq 3,$ $m/2 \geq r\geq 1$ are employed, Chen's algorithm consists of only two decoding steps.
In this case, up to $O\(2^{3m-2r}\)$ functions are called concurrently and Chen's algorithm can be executed in constant parallel time provided majority voting also takes constant parallel time.

The authors in~\cite{Hauck12} investigated how far the number of majority votes in Chen's algorithm can be reduced while focusing on information bits. They established upper and lower bounds for the complexity. But an explicit instruction how to construct a decoder is only provided for a few codes. Furthermore, their decoding process depends on the encoding procedure.

In the present paper, we propose a new hard-decision decoding algorithm for all Reed--Muller codes $\RM\(r,m\)$ with $m \geq 3,$ $m/2 \geq r\geq 1.$ Our decoder is easy to design for software and hardware applications. The algorithm decodes all bits, i.e., information and redundancy, without considering the encoding process. Compared to state-of-the-art majority-logic decoders, our algorithm is less complex. In contrast to recursive decoders~\cite{Dumer04,SchnBos95}, our decoder enables massively parallel decoding in constant parallel time.

The paper is organized as follows. Section~\ref{sect01} introduces the notation and preliminaries on Reed--Muller codes. In Section~\ref{sect02}, we revisit Chen's decoding algorithm and analyze its complexity. In Section~\ref{sect03}, we present in detail our new decoding algorithm including proof of correctness, pseudocode, estimation of complexity and an example for $\RM\(2,5\)$. Our algorithm is compared to Chen's algorithm in terms of complexity in Section~\ref{sect04}. The paper concludes in Section~\ref{sect05} with further advantages of our algorithm in comparison to other classes of decoders.
%
%
\section{Notation and Preliminaries}\label{sect01}
The binary Reed--Muller code $\RM\(r,m\)$ is a $\left[n,k,\delta\right]$ code with
\begin{align*}n:=2^m,\quad k:=\sum_{i=0}^r \binom{m}{i},\quad\delta:=2^{m-r}\end{align*}
which guarantees correcting up to $\delta/2-1$ errors. We number the vectors in $\mathbb{Z}_2^m$ in arbitrary order starting from zero. Every position $i\in\left\{0,1,\ldots,n-1\right\}$ in a binary word of length $n$ is identified by $v_i\in \mathbb{Z}_2^m.$
Then, we characterize a set of vectors $S\subseteq\mathbb{Z}_2^m$ by its incidence vector $\chi_S\in\mathbb{Z}_2^{n}.$ The $i$-th position in $\chi_S$ is set to one if and only if $v_i\in S.$

A \emph{$d$-flat} is a $d$-dimensional affine subspace in $\mathbb{Z}_2^m.$ Given $r$, $m$ and the specific ordering, the Reed--Muller code $\RM\(r,m\)$ is generated by all incidence vectors that characterize $d$-flats with $d= m-r$~\cite{MacWilliamsSloane77}. Therefore, we denote by $\RM\(r,m\)$ not one code but a family of equivalent codes depending on the chosen ordering in $\mathbb{Z}_2^m$.

For the rest of the paper, let $m/2 \geq r\geq 1$, $m \geq 3$. Furthermore, we generally assume a codeword $c:=\(c_0,c_1,\ldots,c_{n-1}\)\in \RM\(r,m\)$ was sent through a noisy channel and \begin{align*}
z&:=\(z_0,z_2\ldots,z_{n-1}\)\\
&=\(c_0,c_1,\ldots,c_{n-1}\)+\(e_0,e_1,\ldots,e_{n-1}\)=c+e\in \mathbb{Z}_2^{n}
\end{align*}
was received where at most $\delta/2-1$ errors occurred.

For any vectors $v,w\in\mathbb{Z}_2^n$, let $v\cdot w\in\mathbb{Z}_2$ denote the scalar product \(over $\mathbb{Z}_2$\) of the two vectors $v$ and $w$.
Let $S\subseteq\mathbb{Z}_2^m$ be arbitrary. The scalar product $z\cdot\chi_S$ is called the \emph{check-sum of $S$}. Since
\begin{align*}
z\cdot\chi_S = \sum_{i\in\left\{j\mid v_j\in S\right\}} z_i\in\mathbb{Z}_2,
\end{align*}
it is not necessary to consider all $n$ entries of $z$. To reduce the complexity of computing the check-sum of $S$, we only take into account the $|S|$ entries $z_i$, $i\in \{j\mid v_j\in S\}$.

In the following, we will say that $S$ \emph{possesses $t$ errors} if and only if
\begin{align*}t =\lv\left\{0\leq i\leq n-1\mid e_{i}\neq 0, v_i\in S\right\}\rv.\end{align*}
In particular, we call $S$ \emph{odd} or \emph{even} if $S$ possesses an odd or even number of errors, respectively. Note that $S$ is odd if and only if $e\cdot\chi_S=1$.

The majority function $\mu:~\left\{0,1\right\}^s \rightarrow \left\{0,1\right\}$ is defined as follows:
\begin{align*}
\mu\(x_1,x_2,\ldots,x_s\)&:=\begin{cases}1&\!\text{if }\lv\left\{0\leq i\leq s: x_i\!=\!1\right\}\rv > \left\lfloor s/2\right\rfloor\\0&\!\text{otherwise}\end{cases}
\end{align*}
where $\lfloor x\rfloor$ represents the largest integer not greater than $x$.

\section{Chen's Two-Step Majority-Logic Decoding of Reed--Muller Codes --- Revisited}\label{sect02}
Chen's decoding algorithm~\cite{Chen71,Chen72} corrects in two majority-logic steps all $n$ positions. It operates on flats of dimension $r+1$ or less and performs majority voting.

\subsection{The Idea}
Chen's algorithm takes advantage of the following proposition.
\begin{proposition}\label{27}
Let $S\subset\mathbb{Z}_2^m$ be arbitrary. Suppose there exist $S_1,\ldots,S_{N}\subseteq \mathbb{Z}_2^m$ with $N\geq \delta-2$ which intersect pairwise in $S$, i.e., $S_i\cap S_j=S$ for all $i,j=1,\ldots,N$, $i\neq j$. Then $S$ is odd if and only if more than $N/2$ sets $S_i$ are odd.
\end{proposition}
\begin{IEEEproof}
Suppose $S$ possesses $t$ errors. Beyond these $t$ errors, up to $\delta/2-1-t$ further errors occurred while transmitting the codeword. Therefore, at least $N-\(\delta/2-1-t\)$ sets $S_i$ must possess the same number of errors as $S$, namely $t$ errors. Note that $N-\(\delta/2-1-t\) \geq N/2+t.$
Hence, if $t$ is odd, more than $N/2$ sets $S_i$ are odd. On the other hand, if $t$ is even, at least $N/2$ sets $S_i$ are even and therefore at most $N/2$ sets $S_i$ are odd.
\end{IEEEproof}

According to Proposition~\ref{27}, it can be deduced whether a set $S$ in $\mathbb{Z}_2^m$ is odd or even, once we have this information about $\delta-2$ arbitrary supersets of $S$, intersecting pairwise in $S$. For some sets, namely  $d$-flats with $d\geq r+1,$ this information can be easily gained. Let us consider a $d$-flat $V$ with  $d\geq r+1.$ Then, its incidence vector, $\chi_V,$ is a codeword of $\RM\(m-r-1,m\)$, the dual code of $\RM\(r,m\)$~\cite{MacWilliamsSloane77}. Thus,
\begin{align}\label{01}z\cdot\chi_V = \underbrace{c\cdot\chi_V}_{=0} + e\cdot\chi_V=e\cdot\chi_V.
\end{align}
Hence, $V$ is odd if and only if the check-sum of $V$ equals one.

Reed~\cite{Reed54} proposed an algorithm comprising $r+1$ steps in which Proposition~\ref{27} is applied. Taking into account the check-sums of certain $\(r+1\)$-flats, the algorithm computes in the first step whether certain $r$-flats are even or odd using majority-logic. In each step $\rho=1,2,\ldots,r+1,$ it is iteratively decided whether the $\(r+1-\rho\)$-flats are odd or even. In the final step, the algorithm yields the number of errors in $0$-flats where every $0$-flat corresponds to a single position.

Analyzing Reed's algorithm, Chen noticed that several steps can be omitted. In the case of $m\geq 3,$ $m/2\geq r\geq 1,$ Chen showed that for every position $i = 0,1,\ldots,n-1,$ there exist $\delta-2$ $r$-flats intersecting pairwise in $\left\{v_i\right\}$. In addition, each $r$-flat is the pairwise intersection of $\delta-2$ $\(r+1\)$-flats~\cite{Chen71,Chen72}. This observation is the basis for a two-step majority-logic algorithm to decode all $n$ positions. The first step is identical to the one in Reed's algorithm where the second step deduces the number of errors in $0$-flats directly from the results for $r$-flats.

\subsection{The Algorithm}
Chen's algorithm operates on a set of flats of dimension 0, $r$ and $r+1,$ say $\mathcal{F},$ which meets the following conditions.
\begin{enumerate}
\item $\left\{v_i\right\}\in\mathcal{F}$ for all $i=0,1,\ldots,n-1.$
\item For every $0$-flat $\left\{v\right\}\in\mathcal{F},$ there exist $r$-flats $V_{0},V_{1},\ldots,V_{\delta-3}\in\mathcal{F}$ with $V_i\cap V_j=\left\{v\right\}$ for $i\neq j.$
\item For every $r$-flat $V\in\mathcal{F}$ there exist $\(r+1\)$-flats $W_{0},W_{1},\ldots,W_{\delta-3}\in\mathcal{F}$ with $W_i\cap W_j=V$ for $i\neq j.$
\end{enumerate}
We call a set of flats \emph{admissible} if it satisfies these three conditions. Furthermore, we say \emph{$W_i$ is used for decoding of $V$} and \emph{$V_i$ is used for decoding of $\left\{v\right\}$}, $i=0,1,\ldots,\delta-3.$

By proving the existence of an admissible set in~\cite{Chen71,Chen72}, Chen indicates a strategy how to decode all positions in two steps using majority-logic. 

\begin{proposition}[Chen's Two-Step Decoding Algorithm]\label{05}
Let $i\in\left\{0,1,\ldots, n-1\right\}$ be arbitrary and let $\mathcal{F}$ be an admissible set.
\begin{enumerate}
\item\label{31} An error occurred at position $i$, i.e., $e_i\neq 0$, if and only if more than half of the $r$-flats used for decoding of $\left\{v_i\right\}$ are odd.
\item\label{32} An $r$-flat $V\in\mathcal{F}$ is odd if and only if  more than half of the $\(r+1\)$-flats used for decoding of $V$ are odd.
\end{enumerate}
We recall that an $\(r+1\)$-flat $W$ is odd if and only if $z\cdot\chi_W=1.$
\end{proposition}
The flats used for decoding are labeled as follows. For all $i = 0,1,\ldots,n-1,$ let $\left\{V_{i,0},\ldots,V_{i,\delta-3}\right\}$ be a set of $r$-flats used for decoding of $\left\{v_i\right\}$ and for all $j = 0,1,\ldots,\delta-3,$ let $\left\{W_{i,j,0},\ldots,W_{i,j,\delta-3}\right\}$ be a set of $\(r+1\)$-flats used for decoding of $V_{i,j}.$ The corresponding algorithm consists of four \emph{function levels}.
\begin{algorithmic}[1]
\Statex
\Input the received word $z\in\mathbb{Z}_2^{n}$
\Require at most $\delta/2-1$ errors occurred
\Output the actual transmitted codeword from $\RM\(r,m\)$
\State $\forall~i = 0,1,\ldots,n-1,~\forall~j,l = 0,1,\ldots,\delta-3$
\Statex $\varsigma_{i,j,l}:= z\cdot\chi_{W_{i,j,l}},$\label{33}
\State $\forall~i = 0,1,\ldots,n-1,~\forall~j = 0,1,\ldots,\delta-3$
\Statex $\mu_{i,j}:=\mu\(\varsigma_{i,j,0},\varsigma_{i,j,1},\ldots,\varsigma_{i,j,\delta-3}\),$\label{29}
\State $\forall~ i = 0,1,\ldots,n-1$
\Statex $\eta_{i}:=\mu\(\mu_{i,0},\mu_{i,1},\ldots,\mu_{i,\delta-3}\),$\label{30}
\State \Return $z+\eta:=\(z_0+\eta_0,\ldots,z_{n-1}+\eta_{n-1}\).$
\end{algorithmic}
The symbol ``$+$'' represents an addition in $\mathbb{Z}_2.$ If not more than $\delta/2-1$ errors occurred, $\eta$ equals the error pattern $e$ such that the actual transmitted codeword $c$ is returned. The term two-step decoding refers to the two steps in line~\ref{29} and~\ref{30} testing for majority.

%
\subsection{The Complexity}
At each of the four function levels, a specific function is called multiple times. All function calls at the same function level can be carried out simultaneously. In Table~\ref{table05}, we specify for each function level how often the corresponding function is called (simultaneously) and how many inputs the function gets.
In total, $O\(n\delta^2\)$ functions are called in Chen's algorithm.

\begin{table}[t]
\setlength{\extrarowheight}{2pt}
\caption{Number of parallel function calls at each function level of Chen's decoding algorithm}
\label{table05}
\centering
\begin{tabular}{|c|c|c|c|}
\firsthline
Function & \multirow{2}{*}{Function}& \multirow{2}{*}{Inputs}& Number of (Parallel)\\
Level && &Function Calls\\\hline\hline
1&Check-Sum &$\(2n/\delta\)$& $n\cdot \(\delta-2\)^2$\\\hline
2&Majority Vote& $\(\delta-2\)$ &$n\(\delta-2\)$	\\\hline
3&Majority Vote& $\(\delta-2\)$& $n$	\\\hline
4&XOR & 2& $n$\\\lasthline
\end{tabular}
\end{table}

\section{Improved Decoding Algorithm}\label{sect03}
Our new decoding algorithm consists of two majority-logic steps. In contrast to Chen, we test less times for majority and compute less check-sums. More precisely, we substitute Step \ref{32} in Chen's decoding procedure \(Proposition~\ref{05}\) by a more efficient method, while we maintain Step \ref{31}. There are two main reasons why our new algorithm is less complex than Chen's decoding procedure. First, instead of considering arbitrary flats for decoding, we use every $r$-flat for all its $2^r$ positions. Second, we never consider $\(r+1\)$-flats. Instead, we developed a new approach where we focus solely on $r$-flats.

\subsection{The Theoretical Approach}\label{sect06}

We start constructing a set of $r$-flats, $\mathcal{F},$ having the characteristics specified in the following proposition.
\begin{proposition}[Proposition 2.3 in~\cite{Hauck12}]\label{03}
There exist $\delta\cdot\(\delta-2\)$ $r$-flats in $\mathbb{Z}_2^m$ such that the intersection of any two of them has at most size 1 and every $v\in\mathbb{Z}_2^m$ is contained in exactly $\delta-2$ of these
$r$-flats.
\end{proposition}
In the proof of this proposition, the authors of~\cite{Hauck12} verify the existence by demonstrating how to construct such a set of  $r$-flats.
At the very beginning, $\delta-2$ $r$-dimensional subspaces in $\mathbb{Z}_2^m$, say $U_0,\ldots, U_{\delta-3}\subset \mathbb{Z}_2^m$, pairwise intersecting in $\left\{0\right\}$ need to be computed. In fact, for $m=ar+b$, $a,b\in\mathbb{N}$, $b<r$, at least $N:=2^b\cdot\(\frac{2^{ar}-1}{2^r-1}-1\)+1$ such subspaces exist~\cite[ch. 1.1, Corollary~2.4]{EisfeldStorme} and can be constructed as shown in~\cite[ch. 1.1, Lemma 2.2,~Corollary~2.3, Corollary~2.4]{EisfeldStorme}.
Note that $N>2^{\(a-1\)r+b}=\delta$. Then, for all $l=0,1,\ldots,\delta-3$, let $W_l:=\left\{w_{l,0},\ldots,w_{l,\delta-1}\right\}$ be a complementary subspace of $U_l$ such that $U_l\oplus W_l=\mathbb{Z}_2^m$. We can state two facts. First, for every vector $v\in\mathbb{Z}_2^m$ and for every subspace $U_l$, $l=0,1,\ldots,\delta-3,$ there exists an $i\in\left\{0,1,\ldots,\delta-1\right\}$ such that $v\in w_{l,i}+U_l.$
Second, every two $r$-flats have at most one vector in common because the intersection of the underlying subspaces is trivial.
Thus, the set of $r$-flats
\begin{align*}\mathcal{F}:=\left\{w_{l,i}+U_l\mid l=0,1,\ldots,\delta-3, i=0,1,\ldots,\delta-1\right\},\end{align*}
comprising $\delta\cdot\(\delta-2\)$ $r$-flats, meets the conditions stated in Proposition~\ref{03}. The algorithm we will propose operates on this set of $r$-flats. Before we present our algorithm, we will explain its mathematical background in Theorem~\ref{07} using the following notations.

\begin{definition}  We define $\(r+1\)$-flats $W_{l,i,j}\subseteq\mathbb{Z}_2^m$ and integers $\varsigma_{l,i}, \mu_l\in \mathbb{Z}_2$
\begin{align*}
W_{l,i,j}&:=w_{l,i}+\langle w_{l,i}+w_{l,j}\rangle \oplus U_l,\\
\varsigma_{l,i}&:= z\cdot\chi_{w_{l,i}+U_l},\\
\mu_l&:=\mu\(\varsigma_{l,0},\varsigma_{l,1},\ldots,\varsigma_{l,\delta-1}\),
\end{align*}
for all $l,i,j,$ with $l=0,1,\ldots,\delta-3,$ $i,j=0,1,\ldots,\delta-1,$ $i\neq j.$
\end{definition}

\begin{theorem}\label{07}\mbox{}
\begin{enumerate}
\item\label{22} An error occurred at position $j\in\left\{0,1,\ldots, n-1\right\}$, i.e., $e_j\neq 0$, if and only if at least $\delta/2$ flats from $\mathcal{F}$ containing $v_j$ are odd.
\item\label{23} A flat $w_{l,i}+U_{l}\in\mathcal{F}$ is odd if and only if $\mu_l\neq\varsigma_{l,i}$.
\end{enumerate}
\end{theorem}

Before we prove Theorem~\ref{07}, we state some general properties of flats.
\begin{lemma}\label{02}Let $l\in\left\{0,1,\ldots,\delta-3\right\}$ be arbitrary. Then
\begin{enumerate}
\item\label{04} $\(w_{l,i}+U_l \)~\cap~\(w_{l,j}+U_l\) = \emptyset,$\label{14}
\item\label{06} $\(w_{l,i}+U_l\)~\dot\cup~ \(w_{l,j}+U_l\) = W_{l,i,j},$\label{15}
\item $W_{l,i,j}~\cap~ W_{l,i,j^{\prime}} = w_{l,i}+U_l,$\label{16}
\end{enumerate}
for all pairwise distinct indices $i,j,j^{\prime}\in\left\{0,\ldots, \delta-1\right\}.$
\end{lemma}
\begin{IEEEproof}
\begin{enumerate}
\item Obvious.
\item Clearly, $\(w_{l,i}+U_l\), \(w_{l,j}+U_l\) \subset W_{l,i,j}$ and with~\ref{04}
\begin{align*}\lv\(w_{l,i}+U_l\)\dot\cup \(w_{l,j}+U_l\)\rv = 2\cdot 2^{r}  = \lv W_{l,i,j}\rv.\end{align*}
\item Follows from~\ref{04} and~\ref{06}.
\end{enumerate}
\end{IEEEproof}

\begin{IEEEproof}[Proof of Theorem \ref{07}]
Assertion~\ref{22} directly follows from Proposition~\ref{27}. Proceeding to part~\ref{23}, let $i\in\left\{0,1,\ldots,\delta-1\right\}$ and $l\in\left\{0,1,\ldots,\delta-3\right\}$ be arbitrary. We will prove that the following statements are equivalent.
\renewcommand{\theenumi}{\roman{enumi}\)}
\renewcommand{\labelenumi}{\theenumi}
\begin{enumerate}
\item\label{37} The flat $w_{l,i}+U_{l}\in\mathcal{F}$ is odd.
\item\label{35} $\lv\left\{0\leq j \leq\delta-1,j\neq i\mid e\cdot \chi_{W_{l,i,j}}=1\right\}\rv\geq\delta/2$.
\item\label{36} $\mu_l \neq \varsigma_{l,i}$.
\end{enumerate}
$\ref{37}\Leftrightarrow\ref{35}$ The $\delta-1$ distinct $\(r+1\)$-flats $W_{l,i,j},$ $j=0,1,\ldots,\delta-1,$ $j\neq i,$ intersect pairwise in $w_{l,i}+U_l$ by Lemma~\ref{02}. Thus, by Proposition~\ref{27}, the flat $w_{l,i}+U_l$ is odd if and only if at least $\delta/2$ of these $\(r+1\)$-flats $W_{l,i,j}$ are odd resulting in the formula stated in~\ref{35}.\\
$\ref{35}\Leftrightarrow\ref{36}$
Similarly to equation~\eqref{01}, we have
\begin{align}
e\cdot \chi_{W_{l,i,j}} &= z\cdot \chi_{W_{l,i,j}} = z\cdot\chi_{W_{l,i,j}}\nonumber\\
&= z\cdot\chi_{w_{l,i}+U_l} + z\cdot\chi_{w_{l,j}+U_l}\nonumber\\
&=\varsigma_{l,i}+\varsigma_{l,j} \label{34}
\end{align}
for every $j=0,1,\ldots,\delta-1,$ $j\neq i,$ where the second equality follows from Lemma~\ref{02}.

We show now that $\lv\left\{0\leq s \leq\delta-1\mid\varsigma_{l,s}=1\right\}\rv \neq \delta/2$. Suppose $\lv\left\{0\leq s \leq\delta-1\mid\varsigma_{l,s}=1\right\}\rv = \delta/2$. It follows from~\eqref{34} that for every $s=0,1,\ldots,\delta-1$ with $\varsigma_{l,s}=1$
\begin{align*}\lv\left\{0\leq j \leq\delta-1,~j\neq s \mid e\cdot \chi_{W_{l,s,j}}=1\right\}\rv=\delta/2.\end{align*}
Applying the already proved equivalence $\ref{37}\Leftrightarrow\ref{35}$, we conclude $w_{l,s}+U_{l}$ is odd for every $s=0,1,\ldots,\delta-1$ with $\varsigma_{l,s}=1$. Thus, $\delta/2$ $r$-flats are odd. Since these $r$-flats are pairwise disjoint by Lemma~\ref{02}~\ref{14}, we have at least $\delta/2$ errors, a contradiction. Hence, $\lv\left\{0\leq s \leq\delta-1\mid\varsigma_{l,s}=1\right\}\rv \neq \delta/2.$

Let us assume $\mu_l \neq \varsigma_{l,i}.$ Then, by the definition of $\mu_l$ and what we have shown before, there exist at least $\delta/2+1$ scalars, say $\varsigma_{l,j_0},\ldots,\varsigma_{l,j_{\delta/2}},$ being unequal to $\varsigma_{l,i}.$ According to equation~\eqref{34}, we have
$e\cdot \chi_{W_{l,i,j_s}} = 1$ for all $s=0,1,\ldots,\delta/2.$

On the other hand, assuming $\mu_l = \varsigma_{l,i}$, there are at most $\delta/2-1$ scalars $\varsigma_{l,j}$ differing from $\varsigma_{l,i}.$ By equation~\eqref{34}, less than $\delta/2$ of the $e\cdot \chi_{W_{l,i,j}}$, $j=0,1,\ldots,\delta-1$, $j\neq i$, are 1.
\renewcommand{\theenumi}{\alph{enumi}\)}
\renewcommand{\labelenumi}{\theenumi}
\end{IEEEproof}

\subsection{The Algorithm}\label{26}
Our new algorithm is strongly based on Theorem~\ref{07}. Tracing back in which $r$-flats every position is contained enables us to design the decoding procedure.

Therefore, we define mappings $\phi_l$, $l=0,1,\ldots,\delta-3$, from $\{0,1,\ldots,n-1\}$ to $\{0,1,\ldots,\delta-1\}$ ensuring that
$v_i\in w_{l,\phi_l(i)}+U_l$ and therefore $v_i+ U_l = w_{l,\phi_l(i)}+U_l$ for all $i=0,1,\ldots,n-1$, $l=0,1,\ldots,\delta-3$.
Once the decoder has been constructed, this mapping between positions and $r$-flats is no longer needed.

\begin{algorithmic}[1]
\Statex
\Input the received word $z\in\mathbb{Z}_2^{n}$
\Require at most $\delta/2-1$ errors occurred
\Output the actual transmitted codeword $c\in\RM\(r,m\)$

\State $\forall~ j=0,1,\ldots,\delta-1,$ $\forall l=0,1,\ldots\delta-3,$\label{39}
\Statex $\varsigma_{l,j}:= z\cdot\chi_{w_{l,j}+U_l},$
\State $\forall~ l=0,1,\ldots,\delta-3$\label{19}
\Statex $\mu_l:= \mu\(\varsigma_{l,0},\varsigma_{l,1},\ldots,\varsigma_{l,\delta-1}\),$

\State $\forall~ i=0,1,\ldots,\delta-1,$ $\forall l=0,1,\ldots,\delta-3$\label{40}
\Statex $\overline{\varsigma_{l,i}}:= \varsigma_{l,i}+\mu_l,$\label{42}

\State $\forall~ j=0,1,\ldots,n-1$\label{21}
\Statex $\eta_j:=\mu\(\overline{\varsigma_{0,\phi_0(j)}}, \overline{\varsigma_{1,\phi_1(j)}},\ldots,\overline{\varsigma_{\delta-3,\phi_{\delta-3}(j)}}\),$

\State \Return $z+\eta:=\(z_0+\eta_0,\ldots,z_{n-1}+\eta_{n-1}\).$\label{43}
\Statex
\end{algorithmic}

First, the scalar $\varsigma_{l,i}$ is computed for every $r$-flat $w_{l,i}+U_l\in\mathcal{F}$. Second, after evaluating the majority function at $\(\varsigma_{l,0},\varsigma_{l,1},\ldots,\varsigma_{l,\delta-1}\)$ for each $l=0,1,\ldots,\delta-3$, the value $\mu_l$ is added to the scalars $\varsigma_{l,0},\ldots,\varsigma_{l,\delta-1}$ where the symbol ``$+$'' represents an addition in $\mathbb{Z}_2$. This guarantees with reference to Theorem~\ref{07} that each $\overline{\varsigma_{l,i}}$ equals one if and only if $w_{l,i}+U_l$ is odd. Finally, the value one is assigned to $\eta_j$ if and only if the majority of the scalars
$\overline{\varsigma_{0,\phi_0(j)}}, \overline{\varsigma_{1,\phi_1(j)}},\ldots,\overline{\varsigma_{\delta-3,\phi_{\delta-3}(j)}}$
assumes one. Provided not more than $\delta/2-1$ errors occurred, $\eta$ equals the error pattern $e$ and $c=z+\eta.$

\subsection{The Complexity}

Our algorithm consists of five function levels. Analogous to Chen's algorithm, a specific function is called multiple times at each function level and all function calls at the same function level can be carried out simultaneously (see Table~\ref{table06}).
Because $m\geq 2r$ and therefore, $\delta^2\geq n$, overall, $O\(\delta^2\)$ functions are called in our algorithm.

\begin{table}[t]
\setlength{\extrarowheight}{2pt}
\caption{Number of parallel function calls at each function level of the proposed decoding algorithm}
\label{table06}
\centering
\begin{tabular}{|c|c|c|c|}
\firsthline
Function &\multirow{2}{*}{Function} & Inputs& Number of (Parallel) \\
Level &&&Function Calls\\\hline\hline
1&Check-Sum&$\(n/\delta\)$ & $\delta\cdot \(\delta-2\)$\\\hline
2&Majority Vote&$\delta$&  $\delta-2$	\\\hline
3&XOR&2 & $\delta\cdot\(\delta-2\)$ \\\hline
4&Majority Vote& $\(\delta-2\)$ & $n$	\\\hline
5&XOR &2& $n$ \\\lasthline
\end{tabular}
\end{table}
%
%
\subsection{An Example for $\RM\(2,5\)$ with Electronic Schematic}
For every $i$, $i=0,1,\ldots,31$, let the vector $v_i:=(v_{i,4},v_{i,3},v_{i,2},v_{i,1},v_{i,0})\in\mathbb{Z}_2^5$ be the binary representation of $i$ such that 
\begin{align*}
i=16\cdot v_{i,4} + 8\cdot v_{i,3} + 4\cdot v_{i,2} + 2\cdot v_{i,1} + v_{i,0}.
\end{align*}
For reasons of clarity, we primarily write $i$ instead of $v_i$.

The Reed--Muller code $\RM\(2,5\)$ is an $[n=32,k=16,\delta=8]$-code correcting three errors. A generator matrix $G$ is given by
\begingroup
\setlength{\arraycolsep}{1.15pt}
\renewcommand{\arraystretch}{1.0}
\begin{align*}\left(
\begin{array}{cccccccccccccccccccccccccccccccc}\\[-0.1em]
1&1&1&1&1&1&1&1&1&1&1&1&1&1&1&1&1&1&1&1&1&1&1&1&1&1&1&1&1&1&1&1\\
0&0&0&0&0&0&0&0&0&0&0&0&0&0&0&0&1&1&1&1&1&1&1&1&1&1&1&1&1&1&1&1\\
0&0&0&0&0&0&0&0&1&1&1&1&1&1&1&1&0&0&0&0&0&0&0&0&1&1&1&1&1&1&1&1\\
0&0&0&0&1&1&1&1&0&0&0&0&1&1&1&1&0&0&0&0&1&1&1&1&0&0&0&0&1&1&1&1\\
0&0&1&1&0&0&1&1&0&0&1&1&0&0&1&1&0&0&1&1&0&0&1&1&0&0&1&1&0&0&1&1\\
0&1&0&1&0&1&0&1&0&1&0&1&0&1&0&1&0&1&0&1&0&1&0&1&0&1&0&1&0&1&0&1\\
0&0&0&0&0&0&0&0&0&0&0&0&0&0&0&0&0&0&0&0&0&0&0&0&1&1&1&1&1&1&1&1\\
0&0&0&0&0&0&0&0&0&0&0&0&0&0&0&0&0&0&0&0&1&1&1&1&0&0&0&0&1&1&1&1\\
0&0&0&0&0&0&0&0&0&0&0&0&0&0&0&0&0&0&1&1&0&0&1&1&0&0&1&1&0&0&1&1\\
0&0&0&0&0&0&0&0&0&0&0&0&0&0&0&0&0&1&0&1&0&1&0&1&0&1&0&1&0&1&0&1\\
0&0&0&0&0&0&0&0&0&0&0&0&1&1&1&1&0&0&0&0&0&0&0&0&0&0&0&0&1&1&1&1\\
0&0&0&0&0&0&0&0&0&0&1&1&0&0&1&1&0&0&0&0&0&0&0&0&0&0&1&1&0&0&1&1\\
0&0&0&0&0&0&0&0&0&1&0&1&0&1&0&1&0&0&0&0&0&0&0&0&0&1&0&1&0&1&0&1\\
0&0&0&0&0&0&1&1&0&0&0&0&0&0&1&1&0&0&0&0&0&0&1&1&0&0&0&0&0&0&1&1\\
0&0&0&0&0&1&0&1&0&0&0&0&0&1&0&1&0&0&0&0&0&1&0&1&0&0&0&0&0&1&0&1\\
0&0&0&1&0&0&0&1&0&0&0&1&0&0&0&1&0&0&0&1&0&0&0&1&0&0&0&1&0&0&0&1\\
\end{array}\right)
\end{align*}
\endgroup

First, the decoder itself needs to be created. As presented in Section~\ref{sect06}, we construct six two-dimensional subspaces $U_0,U_2,\ldots,U_5\subset \mathbb{Z}_2^m$ and corresponding complementary subspaces $W_0,W_2,\ldots,W_5\subset \mathbb{Z}_2^m$,
\begin{flalign*}
U_0&:=\{0, 1, 30, 31\},& W_0&:=\{0, 2, 8, 10, 16, 18, 24, 26\}\\
U_1&:=\{0, 2, 24, 26\},& W_1&:=\{0, 3, 4, 7, 16, 19, 20, 23\}\\
U_2&:=\{0, 3, 20, 23\},& W_2&:=\{0, 4, 8, 12, 18, 22, 26, 30\}\\
U_3&:=\{0, 4, 18, 22\},& W_3&:=\{0, 2, 5, 7, 25, 27, 28, 30\}\\
U_4&:=\{0, 5, 25, 28\},& W_4&:=\{0, 6, 8, 14, 19, 21, 27, 29\}\\
U_5&:=\{0, 6, 27, 29\},& W_5&:=\{0, 1, 8, 9, 22, 30, 23, 31\}
\end{flalign*}
Based on each subspace $U_j$, $j=0,1,\ldots,5$, there exist eight 2-flats $w+U_j$, $w\in W_j.$
\begin{flalign*}
v_{0}+U_0&:=\{0,1,30,31\}, &v_{0}+U_1&:=\{0, 2, 24, 26\},\\
v_{2}+U_0&:=\{2,3,28,29\}, &v_{3}+U_1&:=\{1,3, 25,27\},\\
v_{8}+U_0 &:= \{8,9,22,23\}, &v_{4}+U_1&:=\{4,6,28,30\},\\
v_{10}+U_0&:=\{10,11,20,21\}, &v_{7}+U_1&:=\{5,7,29,31\},\\
v_{16}+U_0&:=\{14, 15, 16,17\}, &v_{16}+U_1&:=\{8,10,16,18\},\\
v_{18}+U_0&:=\{12, 13, 18, 19\}, &v_{19}+U_1&:=\{9, 11, 17, 19\},\\
v_{24}+U_0&:=\{6,7,24,25\}, &v_{20}+U_1&:=\{12,14,20,22\},\\
v_{26}+U_0&:=\{4,5,26,27\}, &v_{23}+U_1&:=\{13,15, 21,23\}.
\end{flalign*}
\begin{flalign*}
v_{0}+U_2&:=\{0, 3, 20, 23\}, &v_{0}+U_3&:=\{0, 4, 18, 22\},\\
v_{4}+U_2&:=\{4, 7, 16, 19\}, &v_{2}+U_3&:=\{2, 6, 16, 20\},\\
v_{8}+U_2 &:= \{8, 11, 28, 31\}, &v_{5}+U_3&:=\{1,5, 19,23\},\\
v_{12}+U_2&:=\{12, 15, 24, 27\}, &v_{7}+U_3&:=\{3, 7, 17, 21\},\\
v_{18}+U_2&:=\{5, 6, 17, 18\}, &v_{25}+U_3&:=\{11, 15, 25, 29\},\\
v_{22}+U_2&:=\{1, 2, 21, 22\}, &v_{27}+U_3&:=\{9, 13, 27, 31\},\\
v_{26}+U_2&:=\{13, 14, 25, 26\}, &v_{28}+U_3&:=\{10, 14, 24, 28\},\\
v_{30}+U_2&:=\{9, 10, 29, 30\}, &v_{30}+U_3&:=\{8, 12, 26, 30\}.
\end{flalign*}
\begin{flalign*}
v_{0}+U_4&:=\{0, 5, 25, 28\}, &v_{0}+U_5&:=\{0, 6, 27, 29\},\\
v_{6}+U_4&:=\{3, 6, 26, 31\}, &v_{1}+U_5&:=\{1, 7, 26, 28\},\\
v_{8}+U_4 &:= \{8, 13, 17, 20\}, &v_{8}+U_5&:=\{8, 14, 19, 21\},\\
v_{14}+U_4&:=\{11, 14, 18, 23\}, &v_{9}+U_5&:=\{9, 15, 18, 20\},\\
v_{19}+U_4&:=\{10, 15, 19, 22\}, &v_{22}+U_5&:=\{11, 13, 16, 22\},\\
v_{21}+U_4&:=\{9, 16, 12, 21\}, &v_{30}+U_5&:=\{3, 5, 24,30\},\\
v_{27}+U_4&:=\{2, 7, 27, 30\}, &v_{23}+U_5&:=\{10, 12, 17, 23\},\\
v_{29}+U_4&:=\{1, 4, 24, 29\}, &v_{31}+U_5&:=\{2, 4, 25, 31\}.
\end{flalign*}
The mappings $\phi_l$, $l=0,1,\ldots,5$, are specified in Table~\ref{table09} ensuring
$v_i+ U_l = w_{l,\phi_l(i)}+U_l$ for all $i=0,1,\ldots,31$.

\begin{table}[t]
\setlength{\extrarowheight}{2pt}
\caption{Mappings $\phi_0,\phi_1,\ldots,\phi_5$}
\label{table09}
\centering
\begin{tabular}{cl}
\firsthline
$l$& $\left(\phi_l(0), \phi_l(1)\ldots, \phi_l(31)\right)$\\\hline\hline
\multirow{2}{*}{$l=0$}&$\left(0, 0, 1, 1, 7, 7, 6, 6,2, 2, 3, 3, 5, 5, 4, 4,\right.$\\
&$\phantom{{}}\left.4, 4, 5, 5, 3, 3, 2, 2,6, 6, 7, 7, 1, 1, 0, 0\right).$\\\hline
\multirow{2}{*}{$l=1$}&$\left(0, 1, 0, 1, 2, 3, 2, 3,4, 5, 4, 5, 6, 7, 6, 7,\right.$\\
&$\phantom{{}}\left.4, 5, 4, 5, 6, 7, 6, 7,0, 1, 0, 1, 2, 3, 2, 3\right).$\\\hline
\multirow{2}{*}{$l=2$}&$\left(0, 5, 5, 0, 1, 4, 4, 1,2, 7, 7, 2, 3, 6, 6, 3,\right.$\\
&$\phantom{{}}\left.1, 4, 4, 1, 0, 5, 5, 0,3, 6, 6, 3, 2, 7, 7, 2\right).$\\\hline
\multirow{2}{*}{$l=3$}&$\left(0, 2, 1, 3, 0, 2, 1, 3,7, 5, 6, 4, 7, 5, 6, 4,\right.$\\
&$\phantom{{}}\left.1, 3, 0, 2, 1, 3, 0, 2,6, 4, 7, 5, 6, 4, 7, 5\right).$\\\hline
\multirow{2}{*}{$l=4$}&$\left(0, 7, 6, 1, 7, 0, 1, 6,2, 5, 4, 3, 5, 2, 3, 4,\right.$\\
&$\phantom{{}}\left.5, 2, 3, 4, 2, 5, 4, 3,7, 0, 1, 6, 0, 7, 6, 1\right).$\\\hline
\multirow{2}{*}{$l=5$}&$\left(0, 1, 7, 5, 7, 5, 0, 1,2, 3, 6, 4, 6, 4, 2, 3,\right.$\\
&$\phantom{{}}\left.4, 6, 3, 2, 3, 2, 4, 6,5, 7, 1, 0, 1, 0, 5, 7\right).$\\\lasthline
\end{tabular}
\end{table}
After constructing the underlying geometrical structure of our decoder, we consider the following example.

Let $m=\(1,1,1,0,0,0,0,0,0,0,0,1,1,1,0,0\)$ be the message word. Then,
\begingroup
\setlength{\arraycolsep}{1.0pt}
\begin{align*}c&=m\cdot G\\
&=\left(
\begin{array}{cccccccccccccccccccccccccccccccc}
1&1&1&1&1&1&0&0&0&1&1&0&0&1&0&1&0&0&0&0&0&0&1&1&1&0&0&1&1&0&1&0
\end{array}\right)
\end{align*}
\endgroup
is the corresponding codeword from $\RM(2,5)$. Suppose $c$ was sent through a noisy channel and
\begingroup
\setlength{\arraycolsep}{1.0pt}
\begin{align*}z&=\left(
\begin{array}{cccccccccccccccccccccccccccccccc}
0&0&1&1&1&1&0&0&0&1&1&0&0&1&0&1&0&0&0&0&0&0&1&1&1&0&0&1&1&0&1&1
\end{array}\right)
\end{align*}
\endgroup
 was received with errors at positions 0,1 and 31.

The decoding can be performed as stated in Section~\ref{26}.
\begin{algorithmic}[1]
\Input $z$
\State $\(\varsigma_{0,0},\varsigma_{0,1},\ldots,\varsigma_{0,7}\)=\(0,1,1,1,1,1,1,1\),$
\Statex $\(\varsigma_{1,0},\varsigma_{1,1},\ldots,\varsigma_{1,7}\)=\(0,0,1,0,1,1,1,1\),$
\Statex $\(\varsigma_{2,0},\varsigma_{2,1},\ldots,\varsigma_{2,7}\)=\(0,1,0,1,1,0,1,1\),$
\Statex $\(\varsigma_{3,0},\varsigma_{3,1},\ldots,\varsigma_{3,7}\)=\(0,1,0,1,1,0,1,1\),$
\Statex $\(\varsigma_{4,0},\varsigma_{4,1},\ldots,\varsigma_{4,7}\)=\(0,0,1,1,1,1,1,0\),$
\Statex $\(\varsigma_{5,0},\varsigma_{5,1},\ldots,\varsigma_{5,7}\)=\(1,1,0,0,0,0,0,1\),$
\Statex
\State $\mu_0:= \mu\(\varsigma_{0,0},\varsigma_{0,1},\ldots,\varsigma_{0,7}\) = \mu\(0,1,1,1,1,1,1,1\)=1$,
\Statex $\mu_1:= \mu\(\varsigma_{1,0},\varsigma_{1,1},\ldots,\varsigma_{1,7}\) = \mu\(0,0,1,0,1,1,1,1\)=1,$
\Statex $\mu_2:= \mu\(\varsigma_{2,0},\varsigma_{2,1},\ldots,\varsigma_{2,7}\) = \mu\(0,1,0,1,1,0,1,1\)=1,$
\Statex $\mu_3:= \mu\(\varsigma_{3,0},\varsigma_{3,1},\ldots,\varsigma_{3,7}\) = \mu\(0,1,0,1,1,0,1,1\)=1,$
\Statex $\mu_4:= \mu\(\varsigma_{4,0},\varsigma_{4,1},\ldots,\varsigma_{4,7}\) = \mu\(0,0,1,1,1,1,1,0\)=1,$
\Statex $\mu_5:= \mu\(\varsigma_{5,0},\varsigma_{5,1},\ldots,\varsigma_{5,7}\) = \mu\(1,1,0,0,0,0,0,1\)=0,$
\Statex
\State $\(\overline{\varsigma_{0,0}},\overline{\varsigma_{0,1},\ldots},\overline{\varsigma_{0,7}}\)$
\Statex $=\(\varsigma_{0,0},\varsigma_{0,1},\ldots,\varsigma_{0,7}\) + \(\mu_0,\mu_0,\mu_0,\mu_0,\mu_0,\mu_0,\mu_0,\mu_0\)$
\Statex $=\(1,0,0,0,0,0,0,0\),$
\Statex $\(\overline{\varsigma_{1,0}},\overline{\varsigma_{1,1},\ldots},\overline{\varsigma_{1,7}}\)=\(1,1,0,1,0,0,0,0\),$
\Statex $\(\overline{\varsigma_{2,0}},\overline{\varsigma_{2,1},\ldots},\overline{\varsigma_{2,7}}\)=\(1,0,1,0,0,1,0,0\),$
\Statex $\(\overline{\varsigma_{3,0}},\overline{\varsigma_{3,1},\ldots},\overline{\varsigma_{3,7}}\)=\(1,0,1,0,0,1,0,0\),$
\Statex $\(\overline{\varsigma_{4,0}},\overline{\varsigma_{4,1},\ldots},\overline{\varsigma_{4,7}}\)=\(1,1,0,0,0,0,0,1\),$
\Statex $\(\overline{\varsigma_{5,0}},\overline{\varsigma_{5,1},\ldots},\overline{\varsigma_{5,7}}\)=\(1,1,0,0,0,0,0,1\),$
\State $\eta_0:=
\mu\(\overline{\varsigma_{0,\phi_0(0)}}, \overline{\varsigma_{1,\phi_1(0)}},\ldots,\overline{\varsigma_{5,\phi_{5}(0)}}\)
=1,$
\Statex $\eta_1:=
\mu\(1,1,1,1,1,1\)=1,$
\Statex $\eta_2:=
\mu\(0,1,1,0,0,1\)=0,$
\Statex $\eta_3:=
\mu\(0,1,1,0,1,0\)=0,$
\Statex ...
\Statex $\eta_{30}:=
\mu\(1,0,0,0,0,0\)=0,$
\Statex $\eta_{31}:=
\mu\(1,1,1,1,1,1\)=1,$
\State \Return $z+\eta \overset{!}{=} c.$
\Statex
\end{algorithmic}
\begin{figure}[t]
\centering
\includegraphics[keepaspectratio=true, scale=0.395, clip=true, trim=0mm 0mm 0mm 0mm]{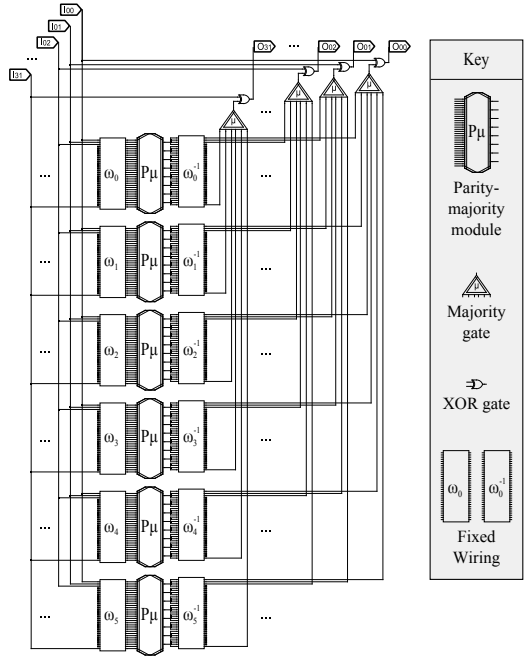}
\caption{The proposed decoder for $\RM\(2,5\)$ with input $z$ and output $c$ provided not more than three errors occurred.}
\label{25}
\end{figure}
\begin{figure}[t]
\centering
\includegraphics[keepaspectratio=true, scale=0.395, clip=true, trim=0mm 0mm 00mm 0mm]{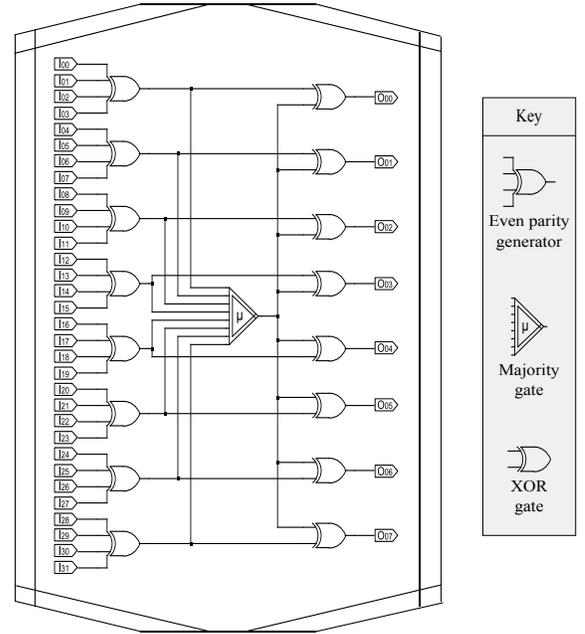}
\caption{The parity-majority module $P\mu$ corresponding to any $l\in\{0,1,\ldots,5\}$ with input $\(z_{\psi_l(0)},z_{\psi_l(1)},\ldots,z_{\psi_l(31)}\)$ and output  $\overline{\varsigma_{l,0}},\overline{\varsigma_{l,1}},
\ldots,\overline{\varsigma_{l,7}}\in\mathbb{Z}_2^{8}.$
In the first layer, even parity generators compute the check-sums and return $\varsigma_{l,0},\varsigma_{l,1},\ldots,\varsigma_{1,7}$ from top to bottom. The majority gate in the second layer returns $\mu_l$. Using XOR gates, $\mu_l$ and $\varsigma_{l,0}, \varsigma_{l,1},\ldots, \varsigma_{l,7}$ are combined in the third layer.}
\label{44}
\end{figure}
\begin{table}[t]
\setlength{\extrarowheight}{2pt}
\caption{Mappings $\psi_0,\psi_1,\ldots,\psi_5$}0
\label{table10}
\centering
\begin{tabular}{cl}
\firsthline
$l$& $\left(\psi_l(0), \psi_l(1)\ldots, \psi_l(31)\right)$\\\hline\hline
\multirow{2}{*}{$l=0$}&$\left(0, 1, 30, 31, 2, 3, 28, 29,8, 9, 22, 23, 10, 11, 20, 21,\right.$\\
&$\phantom{{}}\left.14, 15, 16, 17, 12, 13, 18, 19,6, 7, 24, 25, 4, 5, 26, 27\right).$\\\hline
\multirow{2}{*}{$l=1$}&$\left(0, 2, 24, 26, 1, 3, 25, 27,4, 6, 28, 30, 5, 7, 29, 31,\right.$\\
&$\phantom{{}}\left.8, 10, 16, 18, 9, 11, 17, 19,12, 14, 20, 22, 13, 15, 21, 23\right).$\\\hline
\multirow{2}{*}{$l=2$}&$\left(0, 3, 20, 23, 4, 7, 16, 19,8, 11, 28, 31, 12, 15, 24, 27,\right.$\\
&$\phantom{{}}\left.5, 6, 17, 18, 1, 2, 21, 22,13, 14, 25, 26, 9, 10, 29, 30\right).$\\\hline
\multirow{2}{*}{$l=3$}&$\left(0, 4, 18, 22, 2, 6, 16, 20,1, 5, 19, 23, 3, 7, 17, 21,\right.$\\
&$\phantom{{}}\left.11, 15, 25, 29, 9, 13, 27, 31,10, 14, 24, 28, 8, 12, 26, 30\right).$\\\hline
\multirow{2}{*}{$l=4$}&$\left(0, 5, 25, 28, 3, 6, 26, 31,8, 13, 17, 20, 11, 14, 18, 23,\right.$\\
&$\phantom{{}}\left.10, 15, 19, 22, 9, 16, 12, 21, 2, 7, 27, 30, 1, 4, 24, 29\right).$\\\hline
\multirow{2}{*}{$l=5$}&$\left(0, 6, 27, 29, 1, 7, 26, 28,8, 14, 19, 21, 9, 15, 18, 20,\right.$\\
&$\phantom{{}}\left.11, 13, 16, 22, 3, 5, 24, 30,10, 12, 17, 23, 2, 4, 25, 31\right).$\\\lasthline
\end{tabular}
\end{table}
Fig.~\ref{25} and Fig.~\ref{44} show how the decoding architecture can be built in hardware for a Reed--Muller code $\RM\(2,5\).$
For reasons of clarity and comprehensibility, we structure the decoder (see Fig.~\ref{25}) such that six identical modules, one for every $l=0,1,\ldots,5$, execute line 1, line 2 and line 3 of the proposed algorithm (cf. Section~\ref{26}). A schematic of such a \emph{parity-majority module}, denoted by $P\mu$, is presented in Fig.~\ref{44}.
The blocks labeled with $\omega_0,\omega_1,\ldots,\omega_5$ and $\omega_0^{-1},\omega_1^{-1},\ldots,\omega_5^{-1}$ do not contain any logic gate. They just represent fixed wirings permuting the 32 inputs. The corresponding permutations $\psi_0,\psi_1,\ldots,\psi_5$ are specified in Table~\ref{table10}.

More precisely, within the block $\omega_l$, the input signals, indexed from 0 to 31, are rearranged in the order $\psi_l(0),\psi_l(1),\ldots,\psi_l(31)$ such that the $i$-th signal comes on position $j$ where $\psi_l(j)=i$. Thus, the 32-bit input of the $l$-th module $P\mu$ is just
$\left(z_{\psi_l(0)},z_{\psi_l(1)},\ldots,z_{\psi_l(31)}\right)$.
The module $P\mu$ processes these signals and returns the eight output signals $\left(\overline{\varsigma_{l,0}},\overline{\varsigma_{l,1}},\ldots,\overline{\varsigma_{l,7}}
\right)$.
Recalling that $\overline{\varsigma_{l,i}}$, $l=0,1,,\ldots,5$, $i=0,1,\ldots,7$, states whether $w_{l,i}+U_l$ is odd or even,
every signal $\overline{\varsigma_{l,i}}$ needs to be conveyed to those four different majority gates corresponding to the four vectors contained in $w_{l,i}+U_l$. Therefore, within block $\omega_l^{-1}$, the 32 signals are reordered such that the signal on position $i$, $i=0,1,\ldots,31$, is transferred to position $\psi_l(i)$. Applying this second permutation, it is ensured that the $i$-th signal yields information for determining the $i$-th entry of the codeword,~$c_i$.
%
%

\section{Comparison of Complexity}\label{sect04}
In this section, we compare our algorithm with Chen's algorithm in terms of number of function calls as well as in terms of depth and size of circuits realizing the algorithms. Clearly, the number of function calls is correlated with time complexity where depth and size of a circuit provide information about parallel time and space consumption, respectively.
\subsection{Number of Function Calls}
An overview of the executed functions with respect to the number of inputs and how often each is called in Chen's and the proposed algorithm is provided in Table~\ref{table07}.
\begin{table}[t]
\setlength{\extrarowheight}{2pt}
\caption{Number of function calls in Chen's and the proposed algorithm}
\label{table07}
\centering
\begin{tabular}{|r|r||c|c|}
\firsthline
\multirow{2}{*}{Function}&\multirow{2}{*}{Inputs}&\multicolumn{2}{c|}{Number of Function Calls in}\\\cline{3-4}
&& Chen's Algorithm & the New Algorithm\\\hline\hline
\multirow{2}{*}{Check-Sum} & $\(2n/\delta\)$ & $n\cdot \(\delta-2\)^2$ &0  \\
&$\(n/\delta\)$ & 0 &$\delta\cdot\(\delta-2\)$  \\\hline
\multirow{2}{*}{Majority Vote} & $\delta$&   0& $\delta-2$	\\
&$\(\delta-2\)$&  $n\(\delta-1\)$ & $n$\\\hline
XOR &2&$n$& $n + \delta\cdot\(\delta-2\)$ \\\hline\hline
In Total & & $O\(n\delta^2\)$ & $O\(\delta^2\)$
\\\lasthline
\end{tabular}
\end{table}
\begin{table*}
\setlength{\extrarowheight}{2pt}
\caption{Number of function calls in Chen's and the proposed algorithm for selected Reed--Muller codes}
\label{table08}
\centering
\begin{tabular}{|r||c|c|c||c|c|c||c|c|c||c|c|c|}
\firsthline
\multirow{4}{*}{Function}
&\multicolumn{3}{c||}{RM(2,4)}&\multicolumn{3}{c||}{RM(2,5)}
&\multicolumn{3}{c||}{RM(3,6)}&\multicolumn{3}{c|}{RM(3,7)}
\\\cline{2-13}
&\multirow{3}{*}{Inputs}&\multicolumn{2}{c||}{\# Function Calls in}&\multirow{3}{*}{Inputs}&\multicolumn{2}{c||}{\# Function Calls in}
&\multirow{3}{*}{Inputs}&\multicolumn{2}{c||}{\# Function Calls in}&\multirow{3}{*}{Inputs}&\multicolumn{2}{c|}{\# Function Calls in}
\\
& & Chen's & the New & & Chen's & the New & & Chen's & the New & & Chen's & the New
\\
&& \multicolumn{2}{c||}{Algorithm}
&& \multicolumn{2}{c||}{Algorithm}
&& \multicolumn{2}{c||}{Algorithm}
&& \multicolumn{2}{c|}{Algorithm}
\\\hline\hline
\multirow{2}{*}{Check-Sum}
&8& 64 &0
&8& 1,152 &0
&16& 2,304 &0
&16& 25,088 &0
\\
&4& 0 &8
&4& 0 &48
&8& 0 &48
&8& 0 &224
\\\hline
\multirow{2}{*}{Majority Vote}
& 4& 0& $2$
& 8& 0& $6$
& 8& 0& 6
& 16& 0& 14
\\
& 2 & 48 & 16
& 6 & 224 & 32
& 6 &448 & 64
& 14 &1,920 & 128
\\\hline
XOR
&2 &16& 24
&2 &32& 80
&2 &64& 112
&2 &128& 352
\\\hline\hline
\vtop{%
  \vskip11pt
\hbox{\includegraphics[keepaspectratio=true,scale=0.38, ]
{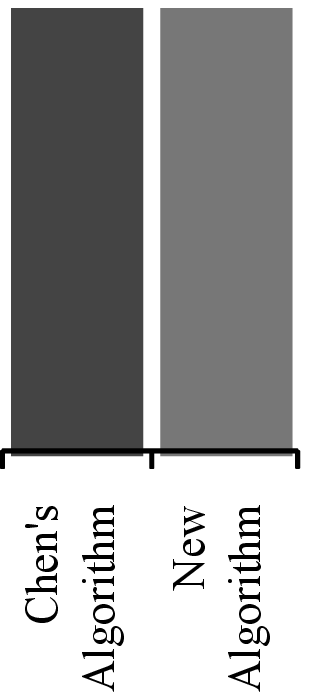}}}
&\multicolumn{3}{c||}{
\vtop{%
  \vskip0pt
  \hbox{%
\includegraphics[keepaspectratio=true,scale=0.38, ]
{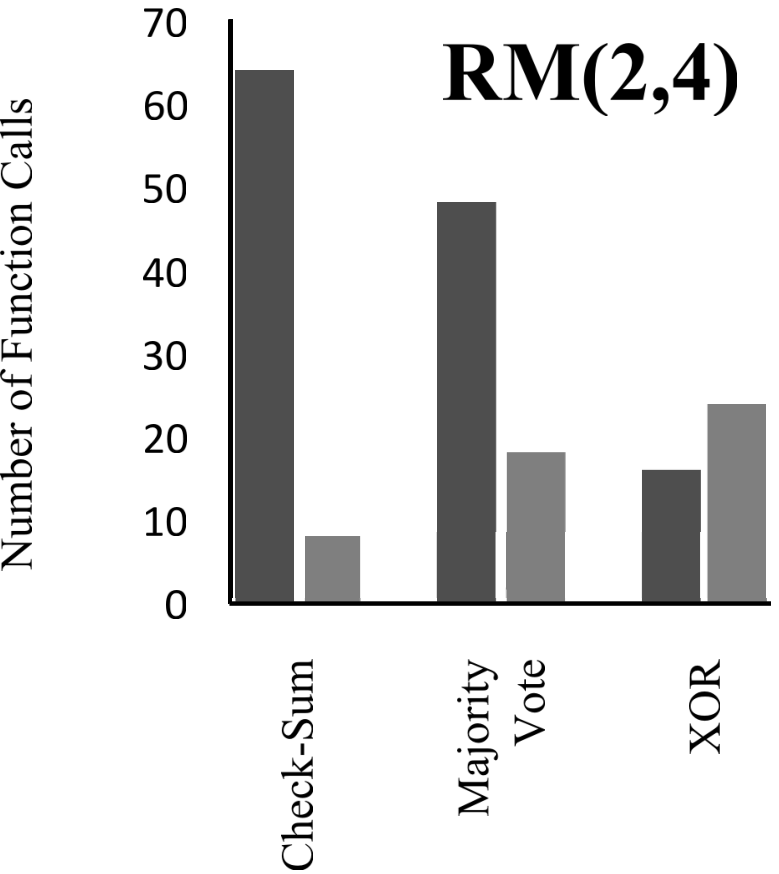}}}}
&\multicolumn{3}{c||}{
\vtop{%
  \vskip0pt
  \hbox{
\includegraphics[keepaspectratio=true,scale=0.38, ]
{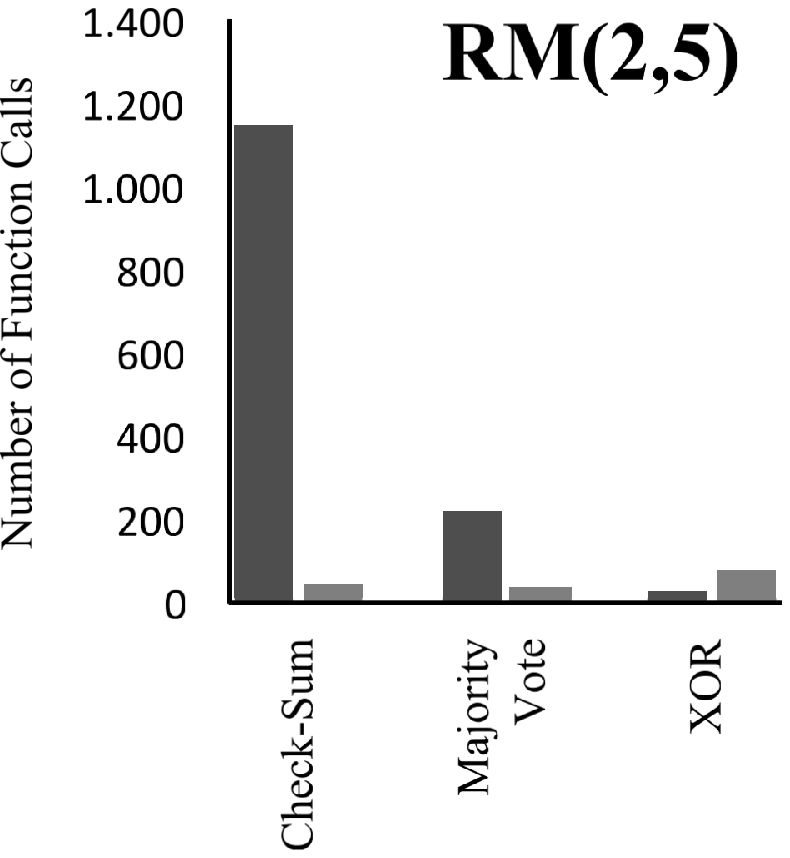}}}}%
&\multicolumn{3}{c||}{
\vtop{%
  \vskip0pt
  \hbox{%
\includegraphics[keepaspectratio=true,scale=0.38, ]
{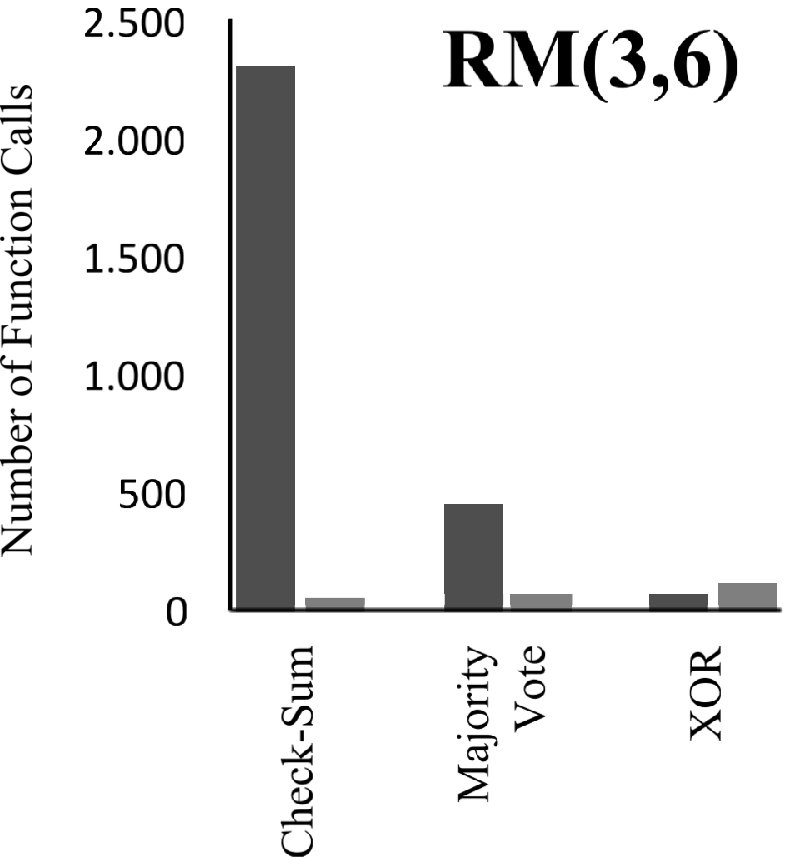}}}}%
&\multicolumn{3}{c|}{
\vtop{%
  \vskip0pt
  \hbox{%
\includegraphics[keepaspectratio=true,scale=0.38, ]
{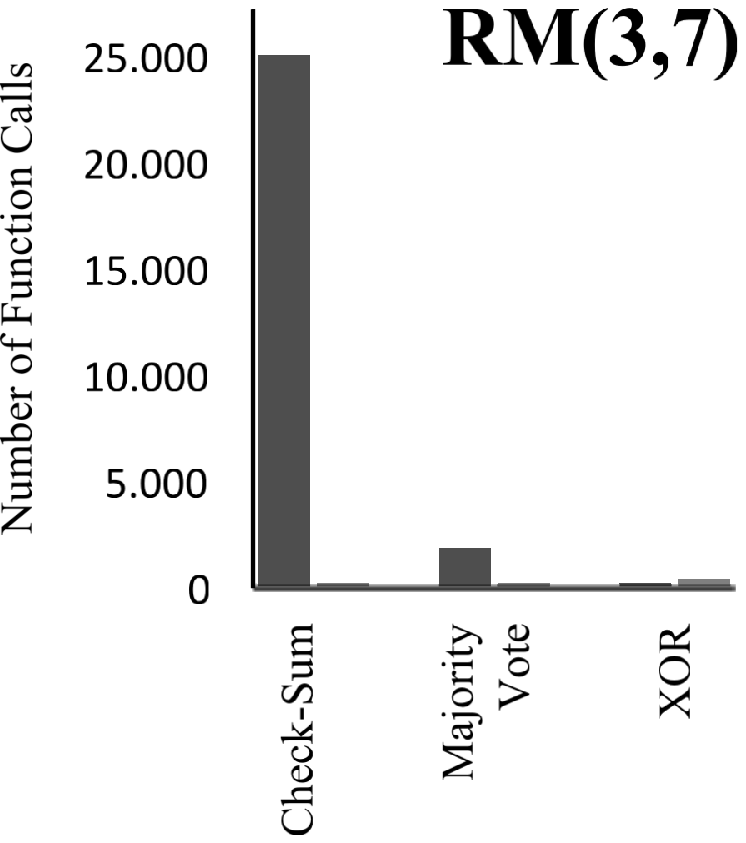}}}}\\\lasthline
\end{tabular}
\end{table*}
Apparently,  decoding with our method instead of Chen's algorithm reduces the number of check-sums to be computed by an order of $n$ and the number of majority votes to be decided by an order of $\delta$. The parameterized data of Table~\ref{table07} is illustrated by way of example in Table~\ref{table08}.
%
%
%
\subsection{Size and Depth of Combinational Circuits}
We want to investigate the size and depth of combinational circuits realizing Chen's and the proposed decoding algorithm. Therefore, we need to consider concrete implementations of the functions, majority vote and check-sum.

In the following, we assume  majority voting is performed in constant time by a single \emph{majority gate}, a specific \emph{linear threshold gate}. Linear threshold gates compute for a given threshold $T\in\mathbb{R}$ and for given weights $w_1,\ldots,w_s\in\mathbb{R}$ the Boolean function $\vartheta: \left\{0,1\right\}^s\rightarrow \left\{0,1\right\}$ where
\begin{align*}\vartheta\(x_1,x_2,\ldots,x_s\):=\begin{cases}
 1& \sum_{i=1}^s w_i\cdot x_i\geq T\\
 0& \text{otherwise}
\end{cases}\end{align*}
\(cf., e.g.,~\cite[ch. 1, sect. 1.1]{Hassoun95}\).
Thus, a majority gate with $s$ inputs is a linear threshold gate where each weight equals one and the threshold equals $\left\lfloor s/2\right\rfloor +1$.

An \emph{even parity generator} is a combinational circuit which computes the \emph{even parity bit} from the input bits. The even parity bit is set to one if and only if the number of input bits which take on the value one is odd. Every check-sum $z\cdot\chi_S$, $S:=\left\{v_{i_1},v_{i_2},\ldots,v_{i_{\lv S\rv}}\right\}\subseteq \mathbb{Z}_2^m$, can be calculated by an even parity generator taking $\left(z_{i_1},z_{i_2},\ldots,z_{i_{\lv S\rv}}\right)$ as input.

Even parity generators of depth $\left\lceil\log_2(N)\right\rceil$ can be simply built out of $N-1$ XOR gates.
It is not surprising that even parity generators with $N$ inputs and of constant depth require more than a polynomial \(in $N$\) number of unbounded fan-in AND, OR and NOT gates~\cite{FurstSaxeSipser81}. But by using linear threshold gates, constant depth and polynomial size can be achieved.\\
Minnick showed in 1961 that an $(2N)$-bit even parity generator of depth 2 can be constructed with $N+1$ linear threshold gates~\cite{Minnick61}. Furthermore, at most $\left\lfloor 2\sqrt{2N-2} + 4\right\rfloor$ linear threshold gates are required for an $(2N)$-bit even parity generator of depth 3~\cite{SRK94}. In fact, the parity function with $N$ inputs can be realized by a threshold circuit of any given depth $d\geq 2$ and size $O\(dN^{1/(d-1)}\)$~\cite{SRK91}.

Recalling the particular function levels of our new algorithm, it can be implemented in a circuit of any given depth $d\geq 6$ and size $s_{\New}(d)=O\(\delta^2\cdot d\cdot{(n/\delta)}^{1/(d-5)}\)$. In this case, the circuit consists of two layers of XOR gates, two layers of majority gates and $d-4$ layers of linear threshold gates. On the other hand, Chen's algorithm can be realized by a circuit of depth $d\geq 5$ and size $s_{\Chen}(d)=O\( n\cdot \delta^2 \cdot d\cdot{(n/\delta)}^{1/(d-4)}\)$ with one layer of XOR gates, two layers of majority gates and $d-3$ layers of linear threshold gates.

Note that for all $d\geq 6$,
\begin{align*}
s_{\Chen}(d)/s_{\New}(d)
= O\(\delta\cdot{(n/\delta)}^{1-1/{((d-4)(d-5))}}\),
\end{align*}
where
\begin{align*}
\min_{d\geq 6}\(\delta\cdot(n/\delta)^{1-1/((d-4)(d-5))}\)=\delta\cdot{(n/\delta)}^{1/2}={(\delta n)}^{1/2}.
\end{align*}

Hence, using our new instead of Chen's algorithm, the size of the decoder with a fixed depth can be reduced by at least an order of $(\delta n)^{1/2}$.
Furthermore, compared to our depth-efficient decoder, the number of gates in a size-efficient circuit realizing Chen's algorithm is still higher by an order of at least $\delta$:
\begin{align*}
s_{\Chen}(d)/s_{\New}(6)= O\( \delta\cdot{(n/\delta)}^{1/(d-4)}\).
\end{align*}
%
%
%
\section{Conclusion}\label{sect05}

In the present paper, we proposed a new hard-decision majority-logic decoding algorithm for Reed--Muller codes $\RM\(r,m\)$ with $m \geq 3,$ $m/2 \geq r\geq 1.$ We showed how to design the decoder by explaining how to construct its underlying geometrical structure. Therefore, our algorithm is easy to implement in both software and hardware. In embedded systems, the proposed decoder can be realized by a simple non-clocked combinational circuit without any registers or flip-flops.
\balance
Regarding the number of operations, recursive decoders~\cite{Dumer04,SchnBos95} usually outperform those based on majority-logic~\cite{Reed54, Chen71, Chen72} and the proposed one.
However, if decoding is to be performed as fast as possible, parallel processing of the functions is appropriate. Clearly, this cannot be sufficiently achieved by recursive algorithms. Their decoding hierarchy is too deeply nested in order to allow fast parallel decoding. Therefore, if algorithms are evaluated on the basis of the required parallel time, majority-logic decoding is preferable to recursive decoding.

We aimed to construct an algorithm which decodes in constant parallel time but is less complex than the best known majority-logic decoders. In fact, Chen's~\cite{Chen71, Chen72} as well as the presented algorithm offers decoding with a constant level of nesting. Nevertheless, using the new method instead of Chen's algorithm, the number of function calls and space consumption can be reduced by at least an order of $n$ and $\delta$, respectively.
Thus, the proposed decoder is a good candidate when massively parallel decoding of all bits in real-time or near real-time is desired.

\section*{Acknowledgment}
The authors thank the anonymous reviewers for their constructive comments and useful suggestions which greatly contributed to improving the manuscript.



\begin{thebibliography}{17}

\bibitem{Reed54}
I.~S. Reed, ``A class of multiple-error-correcting codes and the decoding scheme,''
\emph{IEEE Trans. Inf. Theory}, vol.~4, pp.~38--49, 1954.

\bibitem{Minnick61}
R.~C.~Minnick, ``Linear-input logic,''
\emph{IRE Trans. Electron. Comput.}, vol.EC-10, no.1, pp. 6--16, 1961.

\bibitem{Chen71}
C.-L. Chen, ``On majority-logic decoding of finite geometry codes,''
\emph{IEEE Trans. Inf. Theory}, vol.~17, pp.~332--336, 1971.

\bibitem{Chen72}
C.-L. Chen, ``Note on majority-logic decoding of finite geometry codes,''
\emph{IEEE Trans. Inf. Theory}, vol.~18, pp.~539--541, 1972.

\bibitem{MacWilliamsSloane77}
F.~J.~MacWilliams and N.~J.~A.~Sloane, \emph{The Theory of Error-Correcting Codes}. North-Holland, 1977, 12, impression 2006.

\bibitem{FurstSaxeSipser81}
M.~Furst,  J.~B.~Saxe, and M.~Sipser, ``Parity, circuits, and the polynomial-time hierarchy,''
in \emph{Proc. 1981 IEEE Symp. Found. Comput. Sci}, pp.~260--270.

\bibitem{SRK91}
K.~Siu, V.~Roychowdhury, and T.~Kailath, ``Computing with almost optimal size threshold circuits,''
in \emph{Proc. 1991 IEEE Int. Symp. Inform Theory}, pp. 370, 1991.

\bibitem{SRK94}
K.~Siu, V.~Roychowdhury, and T.~Kailath, ``Depth-size tradeoffs for neural computation,''
\emph{IEEE Trans. Comput. Sci.}, vol.~40, no.~12, pp.~1402--1412, 1991.

\bibitem{Hassoun95}
M. H. Hassoun, \emph{Fundamentals of Artificial Neural Networks}. MIT Press, 1995.

\bibitem{SchnBos95}
G. Schnabl and M. Bossert, ``Soft-decision decoding of Reed--Muller codes as generalized multiple concatenated codes,''
\emph{IEEE Trans. Inf. Theory}, vol.~41, pp.~304--308, 1995.

\bibitem{EisfeldStorme}
J. Eisfeld and L. Storme, ``(Partial) t-spreads and minimal t-covers in finite projective spaces, Lecture notes from the Socrates Intensive Course on Finite Geometry and its Applications, Ghent, Apr. 2000.
Available: http://www.maths.qmul.ac.uk/~leonard/partialspreads/eisfeldstorme.ps.

\bibitem{Blahut03}
R. E. Blahut, ``Codes and algorithms for majority decoding,'' in \emph{Algebraic Codes for Data Transmission} Cambridge University Press, 2003, ch.~13, sec.~3, pp.~430--433.

\bibitem{Dumer04}
I. Dumer, ``Recursive decoding and its performance for low-rate Reed--Muller codes,''
\emph{IEEE Trans. Inf. Theory}, vol.~50, pp.~811--823, 2004.

\bibitem{Freescale11}
Freescale Semiconductor Inc., ``Addressing the challenges of functional safety in the automotive and industrial markets,'' white paper, 2011. Available: http://www.freescale.com/files/microcontrollers/doc/white\_paper/\\FCTNLSFTYWP.PDF.

\bibitem{Issi11}
Integrated Silicon Solution Inc., ``Error correction based 4Mb high speed low power SRAM, 2011. Available: http://www.issi.com/pdf/Highspeed-LowPower-4Mb\_SRAM.PDF.

\bibitem{Huang11}
Q. Huang, S. Lin, and K. A. S. Abdel-Ghaffar,  ``Error-correcting codes for flash coding,''
\emph{IEEE Trans. Inf. Theory}, vol.~57, pp.~6097--6108, 2011.

\bibitem{Hauck12}
P. Hauck, M. Huber, J. Bertram, D. Brauchle, and S. Ziesche, ``Efficient majority-logic decoding of short-length Reed--Muller codes at information positions,'' \emph{IEEE Trans. Commun.}, vol.~61, pp.~930--938, Mar. 2013.

\end{thebibliography}
\end{document}